\documentclass[aps, pra, amsmath, amssymb,atricle, showpacs, superscriptaddress]{revtex4}

\usepackage{graphicx}% Include figure files
\usepackage{bm}% bold math
\begin{document}

\title{Combined error signal in Ramsey spectroscopy of clock transitions}

\author{V. I. Yudin}
\email{viyudin@mail.ru}
\affiliation{Novosibirsk State University, ul. Pirogova 2, Novosibirsk, 630090, Russia}
\affiliation{Institute of Laser Physics SB RAS, pr. Akademika Lavrent'eva 13/3, Novosibirsk, 630090, Russia}
\affiliation{Novosibirsk State Technical University, pr. Karla Marksa 20, Novosibirsk, 630073, Russia}
\affiliation{National Institute of Standards and Technology, Boulder, Colorado 80305, USA}
\author{A. V. Taichenachev}
\affiliation{Novosibirsk State University, ul. Pirogova 2, Novosibirsk, 630090, Russia}
\affiliation{Institute of Laser Physics SB RAS, pr. Akademika Lavrent'eva 13/3, Novosibirsk, 630090, Russia}
\author{M.~Yu.~Basalaev}
\affiliation{Novosibirsk State University, ul. Pirogova 2, Novosibirsk, 630090, Russia}
\affiliation{Institute of Laser Physics SB RAS, pr. Akademika Lavrent'eva 13/3, Novosibirsk, 630090, Russia}
\affiliation{Novosibirsk State Technical University, pr. Karla Marksa 20, Novosibirsk, 630073, Russia}
\author{T.~Zanon-Willette}
\affiliation{Sorbonne Universit$\acute{e}$, Observatoire de Paris, Universit$\acute{e}$ PSL, CNRS, LERMA, F-75005, Paris, France}
\author{T. E. Mehlst$\ddot{{\rm a}}$ubler}
\affiliation{Physikalisch-Technische Bundesanstalt, Bundesallee 100, D-38116 Braunschweig, Germany}
\author{J.~W.~Pollock}
\affiliation{National Institute of Standards and Technology, Boulder, Colorado 80305, USA}
\author{M.~Shuker}
\affiliation{National Institute of Standards and Technology, Boulder, Colorado 80305, USA}
\author{E.~A.~Donley}
\affiliation{National Institute of Standards and Technology, Boulder, Colorado 80305, USA}
\author{J.~Kitching}
\affiliation{National Institute of Standards and Technology, Boulder, Colorado 80305, USA}
%

%\date{\today}

\begin{abstract}
We have developed a universal method to form the reference signal for the stabilization of arbitrary atomic clocks based on Ramsey spectroscopy. Our approach uses an interrogation scheme of the atomic system with two different Ramsey periods and a specially constructed combined error signal (CES) computed by subtracting two error signals with the appropriate calibration factor. CES spectroscopy allows for perfect elimination of  probe-induced light shifts and does not suffer from the effects of relaxation, time-dependent pulse fluctuations and phase-jump modulation errors and other imperfections of the interrogation procedure. The method is simpler than recently developed auto-balanced Ramsey spectroscopy techniques [Ch. Sanner, {\em et al}., Phys. Rev. Lett. \textbf{120}, 053602 (2018); V. I. Yudin, {\em et al}., Phys. Rev. Appl. \textbf{9}, 054034 (2018)], because it uses a single error signal that feeds back on the clock frequency. CES universal technique can be applied to many applications of precision spectroscopy.

\end{abstract}


\maketitle

\subsection*{I. Introduction}
Atomic clocks based on high-precision spectroscopy of isolated quantum systems are currently the most precise scientific instruments, with fractional frequency instabilities and accuracies at the 10$^{-18}$ level \cite{Schioppo_2017, Ludlow_2015,Marti_2018,McGrew_2018,huntemann2016}. Frequency measurements at this level enable improved tests of fundamental physics, as well as new applications like chronometric geodesy \cite{Grotti_2018,Mehlstaubler_2018}.

For many promising clock systems, probe-field-induced frequency shifts can limit the clock frequency instabilities and accuracies.
In the case of magnetically induced spectroscopy \cite{yudin06,bar06}, ac-Stark shifts can limit the achievable clock stability, and for ultranarrow electric octupole \cite{hos09} and two-photon transitions \cite{fis04,badr06}, the large off-resonant ac-Stark shift can completely prevent high-accuracy clock performance. Similarly, the large number of off-resonant laser modes present in clocks based on direct frequency comb spectroscopy \cite{fortier06,stowe08}  induce large ac-Stark shifts. Probe-field-induced shifts also cause instability for microwave atomic clocks based on coherent population trapping (CPT) \cite{Hemmer_JOSAB_1989, Shahriar_1997, Zanon_2005, Pati_2015, Hafiz_2017, Liu_2017}. Compact microwave cold-atom clocks \cite{Esnault_2010,Peng_2015} and hot-cell devices like the POP clock \cite{Micalizio_2012,Godone_2015} that are based on direct microwave interrogation can also be affected by probe-induced frequency shifts.

Probe-induced shifts can be suppressed through the use of Ramsey spectroscopy \cite{rams1950} in combination with cleverly devised modifications. In contrast to continuous-wave spectroscopy, Ramsey spectroscopy has a large number of extra degrees of freedom associated with many parameters that can be precisely controlled: the durations of Ramsey pulses $\tau^{}_1$ and $\tau^{}_2$, the dark time $T$, the phase composition of composite Ramsey pulses \cite{Levitt_1996}, variations in Ramsey sequences including the use of three or more Ramsey pulses, different error signal variants, and so on.
Some modified Ramsey schemes for the suppression of the probe-field-induced shifts in atomic clocks were theoretically described in Ref.~\cite{yudin2010}, which proposed the use of pulses of differing durations ($\tau^{}_1\neq\tau^{}_2$) and the use of composite pulses instead of the standard Ramsey sequence with two equal $\pi/2$-pulses. This ``hyper-Ramsey'' scheme has been successfully realised in an ion clock based on an octupole transition in Yb$^{+}$ \cite{hunt12,huntemann2016}, where a suppression of the light shift by four orders of magnitude and an immunity against its fluctuations were demonstrated. Further developments in Ramsey spectroscopy resulted in additional suppression of probe-field induced frequency shifts. For example, the hyper-Ramsey approach uses new phase variants to construct error signals \cite{NPL_2015, Zanon_2014, Zanon_2016} to  significantly suppress the probe-field-induced shifts in atomic clocks. However, as was shown in Ref.~\cite{Yudin_2016}, all previous hyper-Ramsey methods \cite{yudin2010,hunt12,huntemann2016,NPL_2015,Zanon_2016,Zanon_2015} are sensitive to  decoherence and spontaneous relaxation, which can prevent the achievement of state-of-the-art performance in some systems. To overcome the effect of decoherence, a more complicated construction of the error signal was recently proposed in Ref.~\cite{Zanon_2017}, which requires four measurements for each frequency point (instead of two) combined with the use of the generalized hyper-Ramsey sequences presented in Ref.~\cite{Zanon_2016}. Nevertheless the method in Ref.~\cite{Zanon_2017} is not free from other disadvantages related to technical issues such as time dependent pulse area fluctuations and/or phase-jump modulation errors during the measurements.

The above approaches \cite{yudin2010,hunt12,huntemann2016,NPL_2015,Zanon_2016,Zanon_2015,Zanon_2017} are all one-loop methods, since they use one feedback loop and one error signal. However, frequency stabilization can also be realized with two feedback loops combined with Ramsey sequences with different dark periods $T_1$ and $T_2$ \cite{Yudin_2016,Morgenweg_2014,Sanner_2017}. For example, the synthetic frequency protocol \cite{Yudin_2016} in combination with the original hyper-Ramsey sequence \cite{yudin2010} allows for substantial reduction in the sensitivity to decoherence and imperfections of the interrogation procedure.
Auto-balaced Ramsey spectroscopy  (ABRS) is another effective approach that was first experimentally demonstrated  in a $^{171}$Yb$^+$ ion clock \cite{Sanner_2017}, further substantiated and generalized theoretically in Ref.~\cite{Yudin_2018}, and also recently realized in a CPT atomic clock \cite{Boudot_2018}. For ABRS, in addition to the stabilization of the clock frequency $\omega$, a second loop controls a variable second (concomitant) parameter $\xi$, which is an adjustable property of the first and/or second Ramsey pulses. While both of these two-loop methods \cite{Yudin_2016,Sanner_2017,Yudin_2018} are robust and can perfectly suppress probe-induced shifts of the measurement of the clock frequency, their implementation can be complex due to the two-loop architecture.

A principal question remains: does a one-loop method exist that has comparable (or better) efficiency to ABRS? In this paper, we present a positive answer to this question. We have found a universal protocol to construct a combined error signal (CES), which allows for perfect suppression of probe-induced shifts with the use of only one feedback loop. The CES technique has exceptional robustness, in that it is independent of arbitrary relaxation processes and different non-idealities of the measurement procedure. This method can be considered as a preferred alternative to ABRS spectroscopy. Indeed, CES is technically simpler (because of one feedback loop) and can be more efficient when a hyper-Ramsey pulse sequence  \cite{yudin2010} is used. The CES protocol is applicable to optical atomic clocks as well as to microwave atomic clock based on CPT Ramsey spectroscopy and POP clocks.

\subsection*{II. Theoretical model}

\begin{figure}[t]
\centerline{\scalebox{0.4}{\includegraphics{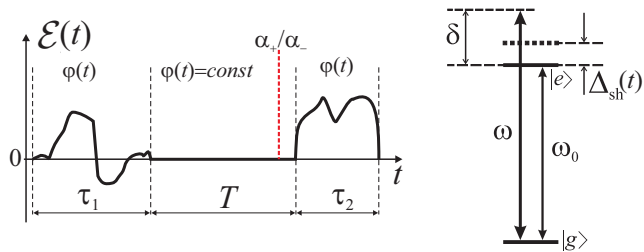}}}\caption{
Left part: schematic illustration of a sequence of two arbitrary Ramsey pulses (with durations $\tau^{}_1$ and $\tau^{}_2$) which are separated by the dark time $T$. Right part: scheme of the clock transition $|g\rangle\leftrightarrow |e\rangle$ (with unperturbed frequency $\omega_0$) interacting with the probe field at the frequency $\omega$.} \label{Fig1}
\end{figure}

We consider a two-level atom with unperturbed frequency $\omega_0$ of the clock transition $|g\rangle\leftrightarrow |e\rangle$ (see Fig.~\ref{Fig1}), which interacts with a Ramsey sequence of two completely arbitrary pulses (with durations $\tau^{}_1$ and $\tau^{}_2$) of the resonant probe field with frequency $\omega$:
\begin{equation}\label{E}
E(t)={\rm Re}\{{\cal E}(t)e^{-i\varphi (t)}e^{-i\omega t}\}\,.
\end{equation}
The pulses are separated by a free evolution interval (dark time) $T$, during which the atom-field interaction is absent (see Fig.~\ref{Fig1}). We emphasise that the Ramsey pulses with arbitrary durations $\tau^{}_1$ and $\tau^{}_2$ can have an arbitrary shape and amplitude (i.e., during $\tau^{}_1$ and $\tau^{}_2$ an amplitude ${\cal E}(t)$ can be an arbitrary real function), and an arbitrary phase function $\varphi(t)$ (e.g., the Ramsey pulses can be composite pulses). In a given sequence of Ramsey measurements, the pulse shape and amplitude must be consistent from one measurement to another. We assume only one restriction: aside from a phase modulation applied to generate the error signal (discussed below), the phase function $\varphi (t)$ should be constant during the dark time $T$, as is typical for Ramsey spectroscopy.

Our main goal is to develop a universal one-loop method, which allows us to stabilize the probe field frequency $\omega$ at the unperturbed frequency of the clock transition, $\omega=\omega_0$, in the presence of decoherence, arbitrary relaxation and light shifts. For this purpose, we will use the formalism of the density matrix $\hat{\rho}$, which has the following form
\begin{equation}\label{rho_din}
\hat{\rho}(t)=\sum_{j,k=g,e}|j\rangle \rho_{jk}^{}(t)\langle k|\,,
\end{equation}
in the basis of states $|g\rangle$ and $|e\rangle$. In the resonance approximation, the density matrix components $\rho_{jk}^{}(t)$ satisfy the following differential equations:
\begin{eqnarray}\label{2_level}
&&[\partial^{}_t+\Gamma-i\tilde{\delta}(t)]\rho^{}_{eg}=i\Omega(t)[\rho^{}_{gg}-\rho^{}_{ee}]/2\,;\quad  \rho^{}_{ge}=\rho^{\ast}_{eg};\nonumber \\
&&[\partial^{}_t+\gamma^{}_{e} ] \rho^{}_{ee}-\gamma^{}_{g\to e} \rho^{}_{gg}=i[\Omega(t)\rho_{ge}-\rho_{eg}\Omega^{\ast}(t)]/2\,,\\
&&[\partial^{}_t+\gamma^{}_{g}]\rho^{}_{gg}-\gamma^{}_{e\to g} \rho^{}_{ee}=-i[\Omega(t)\rho_{ge}-\rho_{eg}\Omega^{\ast}(t)]/2\,.\nonumber
\end{eqnarray}
Here the time dependencies $\Omega(t)$ and $\tilde{\delta}(t)$ are
determined by the following: $\Omega(t)=\langle d\,\rangle {\cal E}(t)e^{-i\varphi (t)}$ and $\tilde{\delta}(t)=\delta-\Delta_{\rm sh}(t)$ during the action of the Ramsey pulses $\tau^{}_1$ and $\tau^{}_2$, but $\Omega(t)=0$ and $\tilde{\delta}(t)=\delta$ during the dark time $T$. $\langle d\,\rangle$ is a matrix element of the atomic dipole moment, $\delta=\omega-\omega_0$ is the detuning of the probe field from the unperturbed atomic frequency $\omega_0$, and $\Delta_{\rm sh}(t)$ is an actual probe-field-induced shift  (see Fig.~\ref{Fig1}) of the clock transition during the Ramsey pulses (e.g., it can be the ac-Stark shift). Also Eq.~(\ref{2_level}) contains five relaxation constants, \{$\gamma^{}_{e}$, $\gamma^{}_{e\rightarrow g}$, $\gamma^{}_{g}$, $\gamma^{}_{g\rightarrow e}$, $\Gamma$\}: $\gamma^{}_{e}$ is a decay rate (e.g., spontaneous) of the exited state $|e\rangle$; $\gamma^{}_{e\rightarrow g}$ is a transition rate (e.g., spontaneous) to the ground state $|g\rangle$; $\gamma^{}_{g}$ is a decay rate of the ground state $|g\rangle$ (e.g., due to black-body radiation and/or collisions); $\gamma^{}_{g\rightarrow e}$ is a transition rate from the ground state $|g\rangle$ to the exited state $|e\rangle$. Note that $\gamma^{}_{e\rightarrow g}=\gamma^{}_{e}$ and $\gamma^{}_{g\rightarrow e}=\gamma^{}_{g}$ in the case of closed two-level system, while $\gamma^{}_{e\rightarrow g}<\gamma^{}_{e}$ and/or $\gamma^{}_{g\rightarrow e}<\gamma^{}_{g}$ in the case of open system. The constant $\Gamma=(\gamma^{}_{e}+\gamma^{}_{g})/2+\widetilde{\Gamma}$ describes the total rate of decoherence: spontaneous as well as all other processes, which are included in the parameter $\widetilde{\Gamma}$ (e.g., an influence of the nonzero spectral width of the probe field).

Equations~(\ref{2_level}) can be rewritten in the vector form
\begin{equation}\label{vect_form}
\partial^{}_t \vec{\rho}(t)=\hat{L}(t)\vec{\rho}(t)\,,
\end{equation}
where $\vec{\rho}(t)$ is a vector formed by the matrix components $\rho_{jk}(t)$,
\begin{equation}\label{rho_vect}
\vec{\rho}(t)=\left(
             \begin{array}{c}
               \rho_{ee}(t) \\
               \rho_{eg}(t) \\
               \rho_{ge}(t) \\
               \rho_{gg}(t) \\
             \end{array}
           \right),
\end{equation}
and operator (Liouvillian) $\hat{L}(t)$ is $4\times 4$ matrix determined by the coefficients of Eq.~(\ref{2_level}):
\begin{equation}\label{L}
\hat{L}(t)=\left(
             \begin{array}{cccc}
               -\gamma^{}_{e} & -i\Omega^{\ast}(t)/2 & i\Omega (t)/2 & \gamma^{}_{g\to e} \\
               -i\Omega (t)/2 & -\Gamma+i\tilde{\delta}(t) & 0 & i\Omega (t)/2 \\
               i\Omega^{\ast}(t)/2 & 0 & -\Gamma-i\tilde{\delta}(t) & -i\Omega^{\ast}(t)/2 \\
               \gamma^{}_{e\to g} & i\Omega^{\ast}(t)/2 & -i\Omega (t)/2 & -\gamma^{}_{g} \\
             \end{array}
           \right).
\end{equation}
In this case, a spectroscopic Ramsey signal can be presented in the following general form, which describes Ramsey fringes (as a function of $\delta$),
\begin{equation}\label{A_Rams}
A_{T}(\delta)=(\vec{\rho}_{\rm obs}^{},\hat{W}^{}_{\tau^{}_2}\hat{G}^{}_T \hat{W}^{}_{\tau^{}_1}\vec{\rho}_{\rm in}^{})\,,
\end{equation}
where the scalar product is determined in the ordinary way: $(\vec{x},\vec{y})=\sum_{m}x^{*}_{m}y^{}_{m}$. Operators $\hat{W}^{}_{\tau^{}_1}$ and $\hat{W}^{}_{\tau^{}_2}$ describe the evolution of an atom during the first ($\tau^{}_1$) and second ($\tau^{}_2$) Ramsey pulses, respectively, and the operator $\hat{G}^{}_T$ describes free evolution during the dark time $T$. Vectors $\vec{\rho}_{\rm in}^{}$ and $\vec{\rho}_{\rm obs}^{}$ are initial and observed states, respectively. For example, if an atom before the Ramsey sequence was in the ground state $|g\rangle$, and after the Ramsey sequence we detect the atom in the exited state $|e\rangle$, then vectors $\vec{\rho}_{\rm in}^{}$ and $\vec{\rho}_{\rm obs}^{}$ are determined, in accordance with definition (\ref{rho_vect}), as
\begin{equation}\label{in_obs}
\vec{\rho}_{\rm in}^{}=\left(
             \begin{array}{c}
               0 \\
               0 \\
               0 \\
               1 \\
             \end{array}
           \right),\quad
           \vec{\rho}_{\rm obs}^{}=\left(
             \begin{array}{c}
               1 \\
               0 \\
               0 \\
               0 \\
             \end{array}
           \right).
\end{equation}
For stabilization of the frequency $\omega$ we need to form an error signal. In our approach, we use phase jumps $\alpha^{}_{+}$ and $\alpha^{}_{-}$ of the probe field in between the first and second Ramsey pulse (see Fig.~\ref{Fig1}), as was proposed in Ref.~\cite{mor89}. These jumps are described by the operators $\hat{\Phi}^{}_{\alpha^{}_+}$ and $\hat{\Phi}^{}_{\alpha^{}_-}$, respectively. In this case, let us introduce the expression of the Ramsey signal in the presence of the pase jump $\alpha$, described by the operator $\hat{\Phi}^{}_{\alpha}$,
\begin{equation}\label{A_Phi}
A_{T}(\delta,\alpha)=(\vec{\rho}_{\rm obs}^{},\hat{W}^{}_{\tau^{}_2}\hat{\Phi}^{}_{\alpha}\hat{G}^{}_T \hat{W}^{}_{\tau^{}_1}\vec{\rho}_{\rm in}^{})\,.
\end{equation}
As a result, the error signal can be presented as a difference,
\begin{equation}\label{err_gen}
S^{\rm (err)}_T = A_{T}(\delta,\alpha^{}_+)-A_{T}(\delta,\alpha^{}_-) = (\vec{\rho}_{\rm obs}^{},\hat{W}^{}_{\tau^{}_2}\hat{D}^{}_{\Phi}\hat{G}^{}_T \hat{W}^{}_{\tau^{}_1}\vec{\rho}_{\rm in}^{})\,,
\end{equation}
with $\hat{D}^{}_{\Phi} = \hat{\Phi}^{}_{\alpha^{}_+}-\hat{\Phi}^{}_{\alpha^{}_-}$. To maximise the error signal, ${\alpha}_{\pm}=\pm \pi/2$ is typically used. However, in real experiments, we can have $|{\alpha}^{}_{+}|\neq |{\alpha}^{}_{-}|$ due to various technical reasons (e.g., electronics) which will lead to a shift of the stabilised frequency $\omega$ in the case of standard Ramsey spectroscopy. Therefore, here we will consider the general case of arbitrary ${\alpha}^{}_{+}$ and ${\alpha}^{}_{-}$ to demonstrate the robustness of CES technique, where the condition $|{\alpha}^{}_{+}|\neq |{\alpha}^{}_{-}|$ does not lead to a frequency shift.

Next we consider the structure of the following operators: $\hat{G}^{}_T$, $\hat{\Phi}^{}_{\alpha^{}_+}$, $\hat{\Phi}^{}_{\alpha^{}_-}$, and $\hat{D}^{}_{\Phi}$. The operator for the free evolution, $\hat{G}^{}_T$, has the following general matrix form
\begin{equation}\label{GT_gen}
\hat{G}^{}_T=\left(
                 \begin{array}{cccc}
                   G_{11}(T) & 0 & 0 & G_{14}(T) \\
                   0 & e^{-(\Gamma -i\delta )T} & 0 & 0 \\
                   0 & 0 & e^{-(\Gamma +i\delta) T} & 0 \\
                   G_{41}(T) & 0 & 0 & G_{44}(T) \\
                 \end{array}
               \right),
\end{equation}
which corresponds to  Eq.~(\ref{vect_form}), if $\Omega(t)=0$ and $\tilde{\delta}(t)=\delta$ in the Liouvillian (\ref{L}). The matrix elements $G_{11}(T)$, $G_{14}(T)$, $G_{41}(T)$, and $G_{44}(T)$ depend on four relaxation rates: \{$\gamma^{}_{e}$, $\gamma^{}_{e\rightarrow g}$, $\gamma^{}_{g}$, $\gamma^{}_{g\rightarrow e}$\}. In particular, for purely spontaneous relaxation of the exited state $|e\rangle$,  when $\gamma^{}_{g}=\gamma^{}_{g\rightarrow e}=0$, we obtain
\begin{equation}\label{GT_sp}
\hat{G}^{}_T=\left(
                 \begin{array}{cccc}
                   e^{-\gamma^{}_{e}T} & 0 & 0 & 0 \\
                   0 & e^{-(\Gamma -i\delta )T} & 0 & 0 \\
                   0 & 0 & e^{-(\Gamma +i\delta) T} & 0 \\
                   \frac{\gamma^{}_{e\rightarrow g}}{\gamma^{}_{e}}(1-e^{-\gamma^{}_{e}T}) & 0 & 0 & 1 \\
                 \end{array}
               \right).
\end{equation}
Operators for the phase jumps $\hat{\Phi}^{}_{\alpha^{}_+}$ and $\hat{\Phi}^{}_{\alpha^{}_-}$ have the forms
\begin{equation}\label{Phi_pm}
\hat{\Phi}^{}_{\alpha_{\pm}}=\left(
                 \begin{array}{cccc}
                   1 & 0 & 0 & 0 \\
                   0 & e^{i\alpha^{}_{\pm}} & 0 & 0 \\
                   0 & 0 & e^{-i\alpha^{}_{\pm}} & 0 \\
                   0 & 0 & 0 & 1 \\
                 \end{array}
               \right),
\end{equation}
which lead to the following expression for $\hat{D}^{}_{\Phi}$,
\begin{equation}\label{DPhi}
\hat{D}^{}_{\Phi}=\hat{\Phi}^{}_{\alpha^{}_+}-\hat{\Phi}^{}_{\alpha^{}_-}=\left(
                 \begin{array}{cccc}
                   0 & 0 & 0 & 0 \\
                   0 & (e^{i\alpha^{}_{+}}-e^{i\alpha^{}_{-}}) & 0 & 0 \\
                   0 & 0 & (e^{-i\alpha^{}_{+}}-e^{-i\alpha^{}_{-}}) & 0 \\
                   0 & 0 & 0 & 0 \\
                 \end{array}
               \right).
\end{equation}
As a result, taking into account Eq.~(\ref{GT_gen}), we obtain a formula for the matrix product $(\hat{D}^{}_{\Phi}\hat{G}^{}_T)$,
\begin{equation}\label{DV}
\hat{D}^{}_{\Phi}\hat{G}^{}_T=\left(
                 \begin{array}{cccc}
                   0 & 0 & 0 & 0 \\
                   0 & e^{-(\Gamma -i\delta )T}(e^{i\alpha^{}_{+}}-e^{i\alpha^{}_{-}}) & 0 & 0 \\
                   0 & 0 & e^{-(\Gamma +i\delta )T}(e^{-i\alpha^{}_{+}}-e^{-i\alpha^{}_{-}}) & 0 \\
                   0 & 0 & 0 & 0 \\
                 \end{array}
               \right)=e^{-\Gamma T}\hat{\Upsilon}^{}_{\delta T}\,,
\end{equation}
where the matrix $\hat{\Upsilon}^{}_{\delta T}$ is defined as
\begin{equation}
\hat{\Upsilon}^{}_{\delta T}=\left(
                 \begin{array}{cccc}
                   0 & 0 & 0 & 0 \\
                   0 & e^{i\delta T}(e^{i\alpha^{}_{+}}-e^{i\alpha^{}_{-}}) & 0 & 0 \\
                   0 & 0 & e^{-i\delta T}(e^{-i\alpha^{}_{+}}-e^{-i\alpha^{}_{-}}) & 0 \\
                   0 & 0 & 0 & 0 \\
                 \end{array}
               \right).
\end{equation}
Note that
\begin{equation}\label{Ups}
\hat{\Upsilon}^{}_{\delta T =0}=\hat{D}^{}_{\Phi}.
\end{equation}
Thus, the error signal (\ref{err_gen}) can be rewritten in the following form:
\begin{equation}\label{err_new}
S^{\rm (err)}_T(\delta)= e^{-\Gamma T}(\vec{\rho}_{\rm obs}^{},\hat{W}^{}_{\tau^{}_2}\hat{\Upsilon}^{}_{\delta T} \hat{W}^{}_{\tau^{}_1}\vec{\rho}_{\rm in}^{}).
\end{equation}
Note that this result is the same if we apply phase jumps $\alpha_{\pm}$ at any arbitrary point during the dark interval $T$.
It is interesting to note that the expression of the error signal in the presence of relaxation is formally different from the the error signal in the absence of relaxation only due to the scalar multiplier $e^{-\Gamma T}$, which primarily affects the amplitude, but not the overall shape of the error signal. This is one of the main specific properties of the phase jump technique for Ramsey spectroscopy that makes it robust against relaxation. Indeed, for other well-known methods of frequency stabilization, which use a frequency jump technique between alternating total periods of Ramsey interrogation $(\tau^{}_1+T+\tau^{}_2)$, relationship (\ref{err_gen}) does not exist. Thus, the phase jump technique has a fundamental advantage over the frequency jump technique in that it is less sensitive to relaxation. In addition, in the ideal case of $\alpha^{}_{+}=-\alpha^{}_{-}=\alpha$, the error signal (\ref{err_gen}) can be expressed as
\begin{eqnarray}\label{err_alpha}
&& S^{\rm (err)}_T(\delta)=2\,{\rm sin}(\alpha) e^{-\Gamma T}(\vec{\rho}_{\rm obs}^{},\hat{W}^{}_{\tau^{}_2}\hat{\Theta}^{}_{\delta T} \hat{W}^{}_{\tau^{}_1}\vec{\rho}_{\rm in}^{}),
\end{eqnarray}
where the matrix $\hat{\Theta}^{}_{\delta T}$,
\begin{equation}
\hat{\Theta}^{}_{\delta T}=\left(
                 \begin{array}{cccc}
                   0 & 0 & 0 & 0 \\
                   0 & ie^{i\delta T} & 0 & 0 \\
                   0 & 0 & -ie^{-i\delta T} & 0 \\
                   0 & 0 & 0 & 0 \\
                 \end{array}
               \right),
\end{equation}
depends only on $\delta T$.

\section*{III. CES protocol}

In this section we demonstrate the universality and robustness of the the CES technique. We use the Ramsey interrogation of the clock transition for two different, fixed intervals of free evolution $T_1$ and $T_2$, where we have two error signals $S^{\rm (err)}_{T_1}(\delta)$ and $S^{\rm (err)}_{T_2}(\delta)$ described by Eq.~(\ref{err_new}). However, for frequency stabilization we introduce the combined error signal (CES) as the following superposition,
\begin{equation}\label{CES_gen}
S^{\rm (err)}_{\rm CES}(\delta)=S^{\rm (err)}_{T_1}(\delta)-\beta^{}_{\rm cal} S^{\rm (err)}_{T_2}(\delta)\,,
\end{equation}
where a calibration coefficient $\beta^{}_{\rm cal}$ is to account for decay of the Ramsey fringe amplitude and will be defined below. Thus, the shift of the stabilized frequency $\bar{\delta}_{\rm clock}$ is determined as a solution of the equation $S^{\rm (err)}_{\rm CES}(\delta)=0$ in relation to the unknown $\delta$.

In  accordance with Eq.~(\ref{err_new}), the expression (\ref{CES_gen}) can be written in the form
\begin{equation}\label{CES_1}
S^{\rm (err)}_{\rm CES}(\delta)=e^{-\Gamma T_1}\left[(\vec{\rho}_{\rm obs}^{},\hat{W}^{}_{\tau^{}_2}\hat{\Upsilon}^{}_{\delta T_1} \hat{W}^{}_{\tau^{}_1}\vec{\rho}_{\rm in}^{})-
\beta^{}_{\rm cal} e^{\Gamma (T_1-T_2)}(\vec{\rho}_{\rm obs}^{},\hat{W}^{}_{\tau^{}_2}\hat{\Upsilon}^{}_{\delta T_2} \hat{W}^{}_{\tau^{}_1}\vec{\rho}_{\rm in}^{})\right].
\end{equation}
If we assume that
\begin{equation}\label{beta}
\beta^{}_{\rm cal} =e^{-\Gamma (T_1-T_2)}\,,
\end{equation}
then we obtain
\begin{equation}\label{CES_2}
S^{\rm (err)}_{\rm CES}(\delta)=e^{-\Gamma T_1}\left[(\vec{\rho}_{\rm obs}^{},\hat{W}^{}_{\tau^{}_2}\hat{\Upsilon}^{}_{\delta T_1} \hat{W}^{}_{\tau^{}_1}\vec{\rho}_{\rm in}^{})-
(\vec{\rho}_{\rm obs}^{},\hat{W}^{}_{\tau^{}_2}\hat{\Upsilon}^{}_{\delta T_2} \hat{W}^{}_{\tau^{}_1}\vec{\rho}_{\rm in}^{})\right].
\end{equation}
In this case, if we apply $\delta=0$ for operators $\hat{\Upsilon}^{}_{\delta T_1}$ and $\hat{\Upsilon}^{}_{\delta T_2}$, then due to Eq.~(\ref{Ups}) we see that
\begin{equation}\label{delta}
S^{\rm (err)}_{\rm CES}(0)=0\,.
\end{equation}
Thus, we have analytically shown that the CES method always leads to zero field-induced shift of the stabilized frequency $\omega$ in an atomic clock, $\bar{\delta}_{\rm clock}=0$.

From a practical viewpoint, it is most important that the calibration coefficient $\beta^{}_{\rm cal}$ [Eq.~(\ref{beta})] does not depend on the values of the phase jumps $\alpha^{}_{\pm}$ used for error signals, or other parameters (such as: amplitude, shape, duration, phase structure $\varphi (t)$, shift $\Delta_{\rm sh}(t)$, etc.) of the two Ramsey pulses $\tau^{}_1$ and $\tau^{}_2$. Thus, $\beta^{}_{\rm cal}$ can be considered as a phenomenological parameter, which is fixed for given setup (via the relaxation constant $\Gamma$) and for given $T_{1,2}$ (via the difference $T_1-T_2$). In the ideal case with no relaxation ($\Gamma =0$), we obtain $\beta^{}_{\rm cal} =1$ for arbitrary $T_{1,2}$. However, in the general case, the value of $\beta^{}_{\rm cal}$ should be empirically determined before long-term frequency stabilization.

As we see from Eq.~(\ref{CES_2}), to maximize the slope of $S^{\rm (err)}_{\rm CES}(\delta)$ it is necessary to use the condition $T_2\ll T_1$. Formally we can even use $T_2=0$ (with the phase jumps $\alpha_{\pm}$ in the virtual point between pulses $\tau^{}_1$ and $\tau^{}_2$). However, due to technical transient regimes (i.e., in acousto-optic modulators) under switching-off/on of Ramsey pulses in real experiments, we believe that it is necessary to keep some nonzero dark time, $T_2 \neq 0$, which significantly exceeds any various transient times. For example, in the case of magnetically-induced spectroscopy \cite{yudin06,bar06}, the transient processes, associated with switching-off/on of magnetic field, can be relatively slow.

Because of the use of two different dark times $T_1$ and $T_2$, CES has some formal similarity to the two-loop methods in Refs.~\cite{Yudin_2016,Sanner_2017, Yudin_2018}. However, the CES technique requires only one feed-back loop for frequency stabilization.

\begin{figure}[t]
\centerline{\scalebox{0.5}{\includegraphics{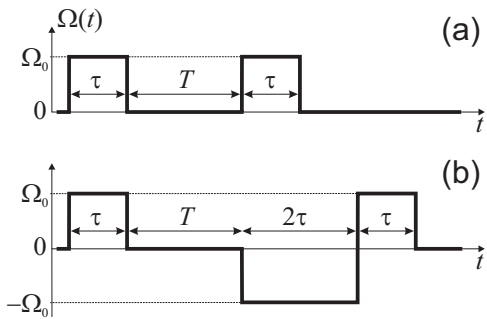}}}\caption{
Two different Ramsey sequences: (a) standard Ramsey sequence \cite{rams1950} with two equal pulses; (b) original hyper-Ramsey sequence \cite{yudin2010} using the composite pulse with phase $\pi$-jump.} \label{R_schemes}
\end{figure}

\subsection*{IV. CES for different Ramsey sequences}

We assume that the main reason for the shift of stabilized frequency $\omega$ arises from probe-induced shift $\Delta_{\rm sh}$ during Ramsey pulses. All calculations are done for ideal case of the phase jumps: $\alpha_+=-\alpha_-=\pi/2$, to maximize the error signal. Also for simplicity, we take into account (for presented calculations) only one relaxation constant $\Gamma$ (rate of decoherence), while all other relaxation constants are negligible: $\gamma^{}_{e}=\gamma^{}_{e\rightarrow g}=\gamma^{}_{g}=\gamma^{}_{g\rightarrow e}=0$, as is typically for high-precision modern atomic clocks based on strongly forbidden optical transition $^1$S$_0$$\rightarrow $$^3$P$_0$ in neutral atoms (such as Mg, Ca, Sr, Yb, Hg) and ions (e.g., Al$^+$, In$^+$), or for octupole transition in the ion Yb$^+$.

In this section, we compare CES spectroscopy for two different pulse sequences: the usual Ramsey sequence with two equal rectangular $\pi/2$-pulses (see Fig.~\ref{R_schemes}a), and the hyper-Ramsey sequence proposed in Ref.~\cite{yudin2010} (see Fig.~\ref{R_schemes}b). If we use the exact calibration coefficient  (\ref{beta}), then both sequences have the identical ideal result, $\bar{\delta}_{\rm clock}= 0$. However, in real experiments, we can know the value of $\beta^{}_{\rm cal}$ with only limited accuracy. In this case, any deviation from the ideal value (\ref{beta}) will lead to the some residual shift of the stabilized frequency, $\bar{\delta}_{\rm clock}\neq 0$, which depends on the type of Ramsey sequence. Thus, there is a problem for the optimal Ramsey sequence with minimal sensitivity to the deviations of $\beta^{}_{\rm cal}$ in Eq.~(\ref{CES_gen}) from the ideal value (\ref{beta}).

Therefore, in our calculations we will use the following expression for calibration coefficient,
\begin{equation}\label{beta_exp}
\beta^{}_{\rm cal} =\chi e^{-\Gamma (T_1-T_2)},
\end{equation}
where the parameter $\chi$ determines the deviation of $\beta^{}_{\rm cal}$ from the ideal value (\ref{beta}). In this case, instead of Eq.~(\ref{CES_2}) we obtain another formula for the CES,
\begin{equation}\label{CES_exp}
S^{\rm (err)}_{\rm CES}(\delta)=e^{-\Gamma T_1}\left[(\vec{\rho}_{\rm obs}^{},\hat{W}^{}_{\tau^{}_2}\hat{\Upsilon}^{}_{\delta T_1} \hat{W}^{}_{\tau^{}_1}\vec{\rho}_{\rm in}^{})-\chi
(\vec{\rho}_{\rm obs}^{},\hat{W}^{}_{\tau^{}_2}\hat{\Upsilon}^{}_{\delta T_2} \hat{W}^{}_{\tau^{}_1}\vec{\rho}_{\rm in}^{})\right],
\end{equation}
where the solution of the equation $S^{\rm (err)}_{\rm CES}(\delta)=0$ (in relation to the unknown $\delta$) determines the residual shift $\bar{\delta}_{\rm clock}$ for the stabilized frequency $\omega$.

\begin{figure}[t]
\centerline{\scalebox{1.0}{\includegraphics{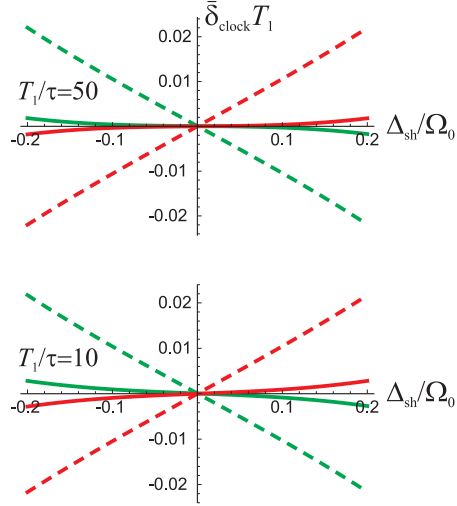}}}\caption{
Comparison of a nonideal CES method for two different pulse sequences: the dashed lines are for a standard Ramsey sequence with two equal pulses (see Fig.~\ref{R_schemes}a), the solid lines are for a hyper-Ramsey sequence using the composite pulse (see Fig.~\ref{R_schemes}b). In the calculations, we assumed in Eq.~(\ref{CES_exp}) a five-percent deviation of $\beta^{}_{\rm cal}$ from the ideal value (\ref{beta}): $\chi =0.95$ (red colored lines) and $\chi =1.05$ (green colored lines), for two different values of $\tau$: $T^{}_1/\tau =50$ (upper figure) and $T^{}_1/\tau =10$ (lower figure). All calculations are done with the following values: $\Omega_0\tau=\pi/2$, $\Gamma=0.5/T^{}_1$, and $T^{}_1/T^{}_2=20$.} \label{compar}
\end{figure}

\begin{figure}[t]
\centerline{\scalebox{1.0}{\includegraphics{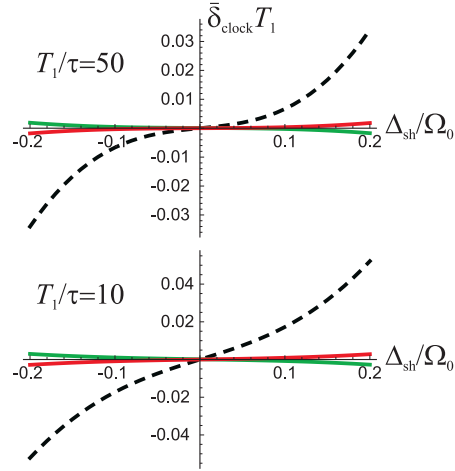}}}\caption{
Comparison of a nonideal CES method, using the hyper-Ramsey sequence (see Fig.~\ref{R_schemes}b), and typical hyper-Ramsey spectroscopy \cite{yudin2010} (black dashed lines, $T=T^{}_1$) for two different values of $\tau$: $T^{}_1/\tau =50$ (upper figure) and $T^{}_1/\tau =10$ (lower figure). In the calculations, we assumed in Eq.~(\ref{CES_exp}) a five-percent deviation of $\beta^{}_{\rm cal}$ from the ideal value (\ref{beta}): $\chi =0.95$ (red solid lines) and $\chi =1.05$  (green solid lines). All calculations are done with the following values: $\Omega_0\tau=\pi/2$, $\Gamma=0.5/T^{}_1$, and $T^{}_1/T^{}_2=20$.} \label{compar_HR}
\end{figure}

In Fig.~\ref{compar} we present a comparison of the CES method for two different pulse sequences: a standard Ramsey sequence with two equal pulses  (see Fig.~\ref{R_schemes}a) and the original hyper-Ramsey sequence \cite{yudin2010} using a composite pulse (see Fig.~\ref{R_schemes}b). In calculations, we have assumed five-percent deviation of $\beta^{}_{\rm cal}$ from the ideal value (\ref{beta}), i.e., $0.95\leqslant\chi\leqslant 1.05$ in Eq.~(\ref{CES_exp}). As we see, the hyper-Ramsey sequence is more robust and persistent, because the use of this scheme leads to a significant reduction of the residual shift $\bar{\delta}_{\rm clock}$ in comparison with the usual Ramsey scheme. In addition, Fig.~\ref{compar_HR} shows that the combination of the CES technique with a hyper-Ramsey sequence significantly exceeds the possibilities of standard hyper-Ramsey spectroscopy \cite{yudin2010}, even for imperfect determination of the calibration coefficient $\beta^{}_{\rm cal}$.

\subsection*{V. Generalized CES and the procedure for frequency stabilization}
The calibration coefficient $\beta^{}_{\rm cal}$ can be estimated as a ratio of the amplitudes of the central Ramsey fringes related to the interrogation procedures with $T_1$ and $T_2$ dark times. However, in this section we describe a more precise method to determine $\beta^{}_{\rm cal}$. For this purpose, we will consider a generalized combined error signal (GCES)
\begin{equation}\label{GCES}
S^{\rm (err)}_{\rm GCES}(\delta)=S^{\rm (err)}_{T_1}(\delta)-\tilde{\beta}(\delta) S^{\rm (err)}_{T_2}(\delta)\,,
\end{equation}
where generalized calibration coefficient $\tilde{\beta}(\delta)$ is a function of $\delta$, which satisfies the following condition,
\begin{equation}\label{beta_GCES}
\tilde{\beta}(0)=\beta^{}_{\rm cal}=e^{-\Gamma (T_1-T_2)}.
\end{equation}
In this case, the stabilized frequency [with the use of GCES (\ref{GCES})] will also always be unshifted, $\bar{\delta}_{\rm clock}=0$.

There are many different variants of the function $\tilde{\beta}(\delta)$. For example, the function $\tilde{\beta}(\delta)$ can be constructed as
\begin{equation}\label{beta12}
\tilde{\beta}(\delta)=\frac{A_{T_1}(\delta,\alpha_+)-A_{T_1}(\delta,\alpha =0)}{A_{T_2}(\delta,\alpha_+)-A_{T_2}(\delta,\alpha =0)}\,;\quad \tilde{\beta}(\delta)=\frac{A_{T_1}(\delta,\alpha_-)-A_{T_1}(\delta,\alpha =0)}{A_{T_2}(\delta,\alpha_-)-A_{T_2}(\delta,\alpha =0)}\,,
\end{equation}
where we use an additional measurement in the absence of phase jump ($\alpha=0$) before the second Ramsey pulse, $A_{T}(\delta,\alpha =0)=(\vec{\rho}_{\rm obs}^{},\hat{W}^{}_{\tau^{}_2}\hat{G}^{}_T \hat{W}^{}_{\tau^{}_1}\vec{\rho}_{\rm in}^{})$. However, another definition,
\begin{equation}\label{beta_symm}
\tilde{\beta}(\delta)=\frac{A_{T_1}(\delta,\alpha_+)+A_{T_1}(\delta,\alpha_-)-2A_{T_1}(\delta,\alpha =0)}{A_{T_2}(\delta,\alpha_+)+A_{T_2}(\delta,\alpha_-)-2A_{T_2}(\delta,\alpha =0)}\,,
\end{equation}
is preferable because of ``symmetry'' in relation to the phase jumps $\alpha^{}_{\pm}$.

\begin{figure}[t]
\centerline{\scalebox{0.7}{\includegraphics{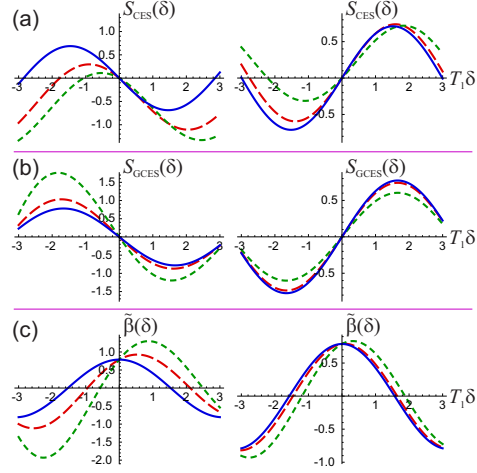}}}\caption{
Comparison of CES and GCES spectroscopies for two different pulse sequences: left panels are for standard Ramsey sequence with two equal pulses (see Fig.~\ref{R_schemes}a), right panels are for hyper-Ramsey sequence (see Fig.~\ref{R_schemes}b). Graphics are presented in the presence of a shift $\Delta_{\rm sh}$ during Ramsey pulses $\tau^{}_1$ and $\tau^{}_2$:  $\Delta_{\rm sh}=0$ (blue solid lines),  $\Delta_{\rm sh}/\Omega_0=0.3$ (red dashed lines), $\Delta_{\rm sh}/\Omega_0=0.5$ (green dashed lines). All calculations are done with the following values: $T^{}_1/\tau =50$, $\Omega_0\tau=\pi/2$, $\Gamma=0.25/T^{}_1$, and $T^{}_1/T^{}_2=20$.\\
(a) signals $S^{\rm (err)}_{\rm CES}(\delta)$ calculated by the use of Eq.~(\ref{CES_gen}) for ideal value of $\beta^{}_{\rm cal}$ [see Eq.~(\ref{beta})];
(b) signals $S^{\rm (err)}_{\rm GCES}(\delta)$  calculated by the use of Eqs.~(\ref{GCES}) and (\ref{beta_symm}) for $\tilde{\beta}(\delta)$;
(c) dependencies $\tilde{\beta}(\delta)$ calculated by the use of Eq.~(\ref{beta_symm}).} \label{err_signals}
\end{figure}

In Fig.~\ref{err_signals}, we compare signals of CES (\ref{CES_gen}) and GCES (\ref{GCES}) for two different pulse sequences (see Fig.~\ref{R_schemes}) in the presence of the field-induced shift $\Delta_{\rm sh}$ (during Ramsey pulses). As we see from Fig.~\ref{err_signals}a, as $\Delta_{\rm sh}$ increases the lineshape $S^{\rm (err)}_{\rm CES}(\delta)$ becomes significantly non-antisymmetrical, while the lineshape $S^{\rm (err)}_{\rm GCES}(\delta)$ (see Fig.~\ref{err_signals}b) maintains its antisymmetry (especially for the hyper-Ramsey scheme, see the right panel in Fig.~\ref{err_signals}b). Fig.~\ref{err_signals}c shows the dependencies of $\tilde{\beta}(\delta)$ calculated by the use of Eq.~(\ref{beta_symm}).

The procedure of frequency stabilization can be organized in conformity with several scenarios. First, we can continually apply GCES (\ref{GCES}) together with Eq.~(\ref{beta_symm}) using six measurements for each frequency point (three different phase jumps, $\alpha =\pm\pi/2,0$, and two different dark times, $T^{}_{1,2}$). However, the use of six measurements can reduce the efficiency of the frequency stabilization, because it increases the length of the interrogation procedure. From our viewpoint, more optimal scenario is the following. In the initial period of frequency stabilization, we use GCES with Eq.~(\ref{beta_symm}). It allows us to determine the calibration coefficient $\beta^{}_{\rm cal}$ [see Eq.~(\ref{beta_GCES})] with satisfactory accuracy, because during measurements we will have the information about the value $\tilde{\beta}(\delta)$ under $\delta\approx 0$. Then the procedure of long-term frequency stabilization can be done with the CES technique (\ref{CES_gen}), using only four measurements for each frequency point (two phase jumps, $\alpha =\pm\pi/2$, and two dark times, $T^{}_{1,2}$). Moreover, we can regularly (but rarely) use GCES again. Indeed, on the one hand, it allows us to do a regular adjustment of the coefficient $\beta^{}_{\rm cal}$ [to eliminate, for example, an influence of possible slow variations of the parameter $\Gamma$ in Eq.~(\ref{beta})]. On the other hand, such intermittent application of GCES will not lead to the significant slowing-down of the process of long-term frequency stabilization.

In addition, as we see from Figs.~\ref{compar}-\ref{err_signals}, the CES or GCES technique works better if the ratio $|\Delta_{\rm sh}/\Omega_0 |$ becomes smaller. Distortions in the error signals arising from this problem can be largely reduced by the use of an additional and well-controllable frequency step $\Delta_{\rm step}$ only during the Ramsey pulses $\tau^{}_1$ and $\tau^{}_2$ \cite{tai09,yudin2010}. In this case, all dependencies presented in Figs.~\ref{err_signals}-\ref{compar_HR} will be the same if we will replace $\Delta_{\rm sh}\rightarrow \Delta_{\rm eff}=(\Delta_{\rm sh}-\Delta_{\rm step})$. Thus, we can always apply a frequency step $\Delta_{\rm step}$ (e.g., with an acousto-optic modulator) during excitation to achieve the condition $|\Delta_{\rm eff}/\Omega_0 | \ll 1$ for an effective shift $\Delta_{\rm eff}$, as it was used in experiments  \cite{hunt12,huntemann2016,NPL_2015,Sanner_2017}.

\subsection*{VI. CES technique for CPT Ramsey spectroscopy}
\begin{figure}[t]
\centerline{\scalebox{0.5}{\includegraphics{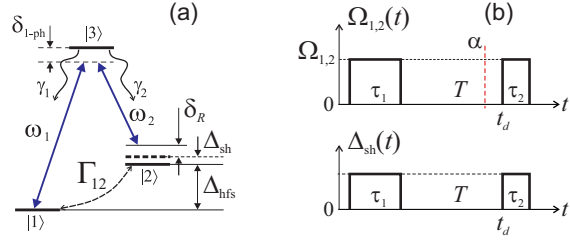}}}\caption{
(a) Atomic three-level $\Lambda$ system. (b) Schematic time dependencies $\Omega_{1, 2}(t)$ and $\Delta_{\text{sh}}(t)$.} \label{Lambda-scheme}
\end{figure}

In this section, we describe the CES technique for Ramsey spectroscopy of the resonances based on coherent population trapping (CPT). As a model, we consider rf CPT resonances that are formed in a three-level $\Lambda$ system under interaction with a resonant bichromatic field,
\begin{equation}\label{bichromatic field}
E(t) = E^{}_{1} e^{-i\omega^{}_{1} t}+E^{}_{2} e^{-i\omega^{}_{2} t}+c.c.\,.
\end{equation}
The CPT resonance is formed when the difference between optical frequencies ($\omega^{}_{1}-\omega^{}_{2}$) is varied near the low-frequency rf transition between lower energy levels $|1\rangle$ and $|2\rangle$: $\omega^{}_{2}-\omega^{}_{1}\approx\Delta_{\text{hfs}}$ [Fig.~\ref{Lambda-scheme}(a)]. In this case, the stabilized rf frequency difference $(\omega^{}_{2}-\omega^{}_{1})$ is the operating frequency for CPT based clocks.

The dynamics of the $\Lambda$ system in the rotating wave approximation are described by the differential equation system for the density matrix components,
\begin{align}\label{rho_eqs}
&[\partial_t+\gamma_\text{opt}-i\delta_\text{1-ph}]\rho^{}_{31}=i\Omega^{}_1(\rho^{}_{11}-\rho^{}_{33})+i\Omega^{}_2\rho^{}_{21}\nonumber \\
&[\partial_t+\gamma_\text{opt}-i\delta_\text{1-ph}]\rho^{}_{32}=i\Omega^{}_2(\rho^{}_{22}-\rho^{}_{33})+i\Omega^{}_1\rho^{}_{12}\nonumber\\
&[\partial_t+\Gamma^{}_{12} -i\delta^{}_R]\rho^{}_{12}=i(\Omega^{\ast}_1\rho^{}_{32}-\rho^{}_{13}\Omega^{}_{2})\\
&[\partial_t+\Gamma^{}_{12} ] \rho^{}_{11}=\gamma^{}_1\rho^{}_{33}+\Gamma^{}_{12} \text{Tr}\{\hat{\rho}\}/2+i(\Omega^{\ast}_1\rho^{}_{31}-\rho^{}_{13}\Omega^{}_{1})\nonumber\\
&[\partial_t+\Gamma^{}_{12} ] \rho^{}_{22}=\gamma^{}_2\rho^{}_{33}+\Gamma^{}_{12} \text{Tr}\{\hat{\rho}\}/2+i(\Omega^{\ast}_2\rho^{}_{32}-\rho^{}_{23}\Omega^{}_{2})\nonumber\\
&[\partial_t+\Gamma^{}_{12} +\gamma] \rho^{}_{33}= i(\Omega^{}_1\rho^{}_{13}-\rho^{}_{31}\Omega^{\ast}_{1})+i(\Omega^{}_2\rho^{}_{23}-\rho^{}_{32}\Omega^{\ast}_{2})\nonumber\\
&\rho^{}_{jk}=\rho^{\ast}_{kj}\;(j,k=1,2,3);\;\; \text{Tr}\{\hat{\rho}\}=\rho^{}_{11}+\rho^{}_{22}+\rho^{}_{33}=1.\nonumber
\end{align}
Here $\delta_\text{1-ph}$ is the one-photon detuning of frequency components $\omega^{}_{1}$ and $\omega^{}_{2}$ from the optical transitions (see Fig.~\ref{Lambda-scheme}); $\delta^{}_R = \omega^{}_{2}-\omega^{}_{1}-\Delta_{\text{hfs}}-\Delta_{\rm sh}(t)$ is the two-photon (Raman) detuning; $\Omega_{1}(t)$$=$$d_{31}E_{1}(t)/$$\hbar$ and $\Omega_{2}(t)$$=$$d_{32}E_{2}(t)/$$\hbar$ are the Rabi frequencies for the transitions $|1\rangle$$\leftrightarrow$$|3\rangle$ and $|2\rangle$$\leftrightarrow$$|3\rangle$ ($d_{31}$ and $d_{32}$ are reduced matrix elements of dipole moment for these transitions); $\gamma$ is the spontaneous decay rate of upper level $|3\rangle$; $\gamma_\text{opt}$ is rate of decoherence (spontaneous, collisional, etc.) of the optical transitions $|1\rangle$$\leftrightarrow$$|3\rangle$ and $|2\rangle$$ \leftrightarrow$$|3\rangle$ (in the case of pure spontaneous relaxation $\gamma_\text{opt}=\gamma/2$); $\gamma^{}_1$ and $\gamma^{}_2$ are corresponding spontaneous decay rates for different channels ($\gamma^{}_1+\gamma^{}_2=\gamma$ in the case of closed $\Lambda$ system); $\Gamma^{}_{12}$ is the relatively slow ($\Gamma^{}_{12}\ll \gamma,\gamma^{}_\text{opt}$) rate of relaxation to the equilibrium isotropic ground state: $\hat{\rho}^{}_0=(|1\rangle\langle 1|+|2\rangle\langle 2|)/2$. Note that $\Delta_\text{sh}(t)$ is an additional actual shift (AC Stark shift) between levels $|1\rangle$ and $|2\rangle$ during the pulses, which results from off-resonant interactions of components of the laser field with different hyperfine states (e.g., Ref.~\cite{Pollock_2018}).

In the case of Ramsey excitation, the scheme of the time dependencies $\Omega_{1}(t)$ and $\Omega_{2}(t)$ is shown in Fig.~\ref{Lambda-scheme}(b), where the first pulse (with duration $\tau^{}_{1}$) prepares an atomic coherence between lower levels $|1\rangle$ and $|2\rangle$, $T$ is the free evolution interval, and the second pulse (with duration $\tau^{}_{2}$) is the detecting pulse, which forms a spectroscopic Ramsey signal. The time dependence $\Delta_{\text{sh}}(t)$ is also shown. If $\tau_{1}$ is much longer than the time for the atoms to enter the dark state, then at the end of first pulse (before the free evolution interval) we have a steady-state condition. In this case, the transient frequency shift, described in \cite{Hemmer_JOSAB_1989}, becomes equal to zero. As a result, the residual shift of the central Ramsey fringe $\bar{\delta}_{\rm clock}=\omega^{}_{2}-\omega^{}_{1}-\Delta_{\text{hfs}}$ results from the off-resonant shift $\Delta_{\text{sh}}$, which is present only during Ramsey pulses ($\tau^{}_{1}$ and $\tau^{}_{2}$) [Fig.~\ref{Lambda-scheme}(b)]. $\Delta_{\text{sh}}$ is the well known AC Stark shift, which is proportional to the total light field intensity $I$.

Instead of Eq.~(\ref{A_Phi}), for calculations of the CPT spectroscopic signal we use the absorption (spontaneous scattering), which is proportional to the integral value during the second pulse $\tau_2$  starting at time $t^{}_{d}$ [Fig.~\ref{Lambda-scheme}(b)],
\begin{equation}\label{A_CPT}
A^{\rm (CPT)}_{T}(\delta,\alpha)=\int_{t^{}_{d}}^{t^{}_{d}+\tau^{}_2}\rho^{}_{33}(t')dt',
\end{equation}
where we have introduced the phase jump $\alpha$ during the dark time $T$ (e.g., Ref.~\cite{Guerandel_2007}). This phase jump describes a phase difference of the product $(E^{}_1E^{\ast}_2)_{\tau^{}_1}$ during the first Ramsey pulse $\tau^{}_1$ and the product $(E^{}_1E^{\ast}_2)_{\tau^{}_2}$ during the second pulse $\tau^{}_2$,
\begin{equation}\label{CPT_phase}
(E^{}_1E^{\ast}_2)_{\tau^{}_2}=e^{-i\alpha}(E^{}_1E^{\ast}_2)_{\tau^{}_1}.
\end{equation}
Using the determination of the signal (\ref{A_CPT}) in formulas (\ref{err_gen}) and ({\ref{CES_gen}})-(\ref{beta_symm}) from the previous sections, we describe a realization of the CES/GCES techniques for CPT Ramsey spectroscopy. In this case, it is necessary to use  $\Gamma^{}_{12}$ instead of $\Gamma$.

\subsection*{Conclusion}
We have developed a universal one-loop method to form the reference signal for stabilization of arbitrary atomic clocks based on Ramsey spectroscopy. This method uses the interrogation of an atomic system for two different Ramsey periods and a specially constructed combined error signal (CES) [see Eq.~(\ref{CES_gen})]. The CES technique requires four measurements for each frequency point as well as a preliminary measurement (or estimation) of the calibration coefficient  $\beta^{}_{\rm cal}$. It was shown that the most robustness is achieved with the combination of the CES protocol and a hyper-Ramsey pulse sequence (see in Ref.~\cite{yudin2010}). Also a method of generalized combined error signal (GCES) was developed [see Eq.~(\ref{GCES})], which requires six measurements for each frequency point and has an exceptional robustness. The CES/GCES spectroscopy allows for perfect elimination of probe-induced light shifts and does not suffer from the effects of relaxation, time-dependent pulse fluctuations and phase-jump modulation errors and other non-idealities of the interrogation procedure. A variant of the frequency stabilization using CES with intermittent GCES protocols has been proposed. In addition, the applicability of CES/GCES techniques for CPT atomic clocks has been described. The implementation of this approach can lead to significant improvement of the accuracy and long-term stability for a variety of types of atomic clocks.

Also, it will be interesting to experimentally compare the one-loop CES/GCES method with the two-loop auto-balanced Ramsey spectroscopy (ABRS) \cite{Sanner_2017,Yudin_2018,Boudot_2018}. We believe that both methods have comparable efficiency of the frequency stabilization, but CES/GCES is more simple technically due to only one feedback loop. Moreover, in the case of optical transitions, the CES/GCES protocol with the use of hyper-Ramsey pulse sequence (see in Ref.~\cite{yudin2010}) can be even more efficient in comparison with ABRS.

\begin{acknowledgments}
We thank E. Ivanov, J. Elgin, C. Oates, M. Lombardi, Ch. Sanner, Ch. Tamm, E. Peik, and N. Huntemann for useful discussions and comments. This work was supported by the Russian Scientific Foundation (No. 16-12-10147).  Contributions to this article by workers at NIST, an agency of the U.S. Government, are not subject to U.S. copyright.
%This work was supported by the Russian
%Scientific Foundation (No. 16-12-10147). M. Yu. Basalaev was also supported by the Ministry of Education and Science of the Russian
%Federation (No. 3.1326.2017/4.6), and Russian Foundation for Basic Research (No. 17-02-00570 and 16-32-60050 mol\_a\_dk).
\end{acknowledgments}

\end{document}